\documentclass[runningheads]{llncs}
\usepackage{graphicx}
\usepackage{amsmath}
\usepackage{amssymb}
\usepackage{textcomp,gensymb}
\usepackage{algorithm}
\usepackage{algpseudocode}
\usepackage{tabularx,booktabs}
\usepackage{pdflscape}
\usepackage{afterpage}
\newcolumntype{Y}{>{\centering\arraybackslash}X}
\newcommand\inlineeqno{\stepcounter{equation}\ (\theequation)}
\newcommand{\beginsupplement}{%
        \setcounter{table}{0}
        \renewcommand{\thetable}{S\arabic{table}}%
        \setcounter{figure}{0}
        \renewcommand{\thefigure}{S\arabic{figure}}%
     }

\begin{document}
\title{Spatio-temporal motion correction and iterative reconstruction of in-utero fetal fMRI}
\titlerunning{Spatio-temporal reconstruction of fetal fMRI}

\author{
Athena Taymourtash\inst{1} \and
Hamza Kebiri\inst{2,3} \and
Ernst Schwartz\inst{1} \and
Karl-Heinz Nenning\inst{1,4} \and
Sébastien Tourbier\inst{2} \and
Gregor Kasprian \inst{5} \and
Daniela Prayer \inst{1,5} \and
Meritxell Bach Cuadra\inst{2,3} \and
Georg Langs\inst{*1}
}
%
%index{Taymourtash, Athena}
%index{Kebiri, Hamza}
%index{Schwartz, Ernst}
%index{Nenning, Karl-Heinz}
%index{Tourbier, Sébastien}
%index{Kasprian, Gregor}
%index{Prayer, Daniela}
%index{Bach Cuadra, Meritxell}
%index{Langs, Georg}

\authorrunning{A. Taymourtash et al.}
\institute{Computational Imaging Research Lab, Department of Biomedical Imaging and Image-guided Therapy, Medical University of Vienna, Austria. \and
Medical Image Analysis Laboratory, Department of Radiology, Lausanne University Hospital and University of Lausanne, Switzerland. \and
CIBM Center for Biomedical Imaging, Switzerland. \and
Center for Biomedical Imaging and Neuromodulation, Nathan Kline Institute, Orangeburg, NY, United States. \and
Division of Neuroradiology and Muskulo-skeletal Radiology, Department of Biomedical Imaging and Image-guided Therapy, Medical University of Vienna, Austria. \\
\email{georg.langs@meduniwien.ac.at}}
\maketitle              % typeset the header of the contribution
\begin{abstract}
Resting-state functional Magnetic Resonance Imaging (fMRI) is a powerful imaging technique for studying functional development of the brain \textit{in utero}. However, unpredictable and excessive movement of fetuses have limited its clinical applicability. Previous studies have focused primarily on the accurate estimation of the motion parameters employing a single step 3D interpolation at each individual time frame  to recover a motion-free 4D fMRI image. Using only information from a 3D spatial neighborhood neglects the temporal structure of fMRI and useful information from neighboring timepoints. Here, we propose a novel technique based on four dimensional iterative reconstruction of the motion scattered fMRI slices. Quantitative evaluation of the proposed method on a cohort of real clinical fetal fMRI data indicates improvement of reconstruction quality compared to the conventional 3D interpolation approaches.

\keywords{Fetal fMRI \and image reconstruction \and motion-compensated recovery \and regularization.}
\end{abstract}
\section{Introduction}
Functional magnetic resonance imaging (fMRI) offers a unique means of observing the functional brain architecture and its variation during development, aging, or disease.  Despite the insights into network formation and functional growth of the brain,\textit{in utero} fMRI of living human fetuses, and the developmental functional connectivity (FC), however, remain challenging.
%mainly because of the uncontrollable excessive fetal movement. 
Since the fMRI acuisition takes several minutes, unconstrained and potentially large movements of the fetuses, uterine contractions, and maternal respiration can cause severe artifacts such as in-plane blurring, slice cross-talk, and spin-history artifacts that likely vary over time. Without mitigation, motion artifacts can considerably affect the image quality, leading to a bias of subsequent conclusions about the FC of the developing brain.

Standard motion correction approaches, including frame-by-frame spatial realignment 
%regression of motion estimates, and filtering out motion components using independent components analysis 
along with discarding parts of data with excessive motion, have been adopted so far to address motion artifacts of \textit{in utero} fMRI \cite{you2016robust,rutherford2019observing,turk2017spatiotemporal}. 
%However, the reported rejection rate of fMRI frames in these studies is very high and many subjects were removed since there was not enough data left after censoring. Further attempts to correct for fetal fMRI motion-related artifacts mainly relied on improving the estimate of the realignment parameters for potentially large movement of the fetuses. 
More recently, cascaded slice-to-volume registration\,\cite{seshamani2013cascaded} combined with spin history correction\,\cite{ferrazzi2014resting}, and framewise registration based on the 2\textsuperscript{nd} order edge features instead of raw intensities\,\cite{scheinost2018fetal} were suggested. These studies used 3D linear interpolation of motion scattered data at each volume independently to reconstruct the entire time series. 
%
%Since \textit{in utero} motion is unconstrained and complex, the rate of motion scattered points versus the regular points to be infered is less than 1 to 1 in each 3D volume. Therefore, when strong motion occurs, gaps might be open up in the interpolated image since there would be no data contributing to the reconstruction of a given regular grid. 
%
Since \textit{in utero} motion is unconstrained and complex, the regular grid of observed fMRI volumes becomes a set of irregularly motion scattered points possibly out of the field-of-view of the reconstruction grid, which might contain gaps in regions with no points in close proximity. Therefore interpolation in each 3D volume cannot recover the entire reconstruction grid. 
%

\begin{figure}[t]
\includegraphics[width=\textwidth]{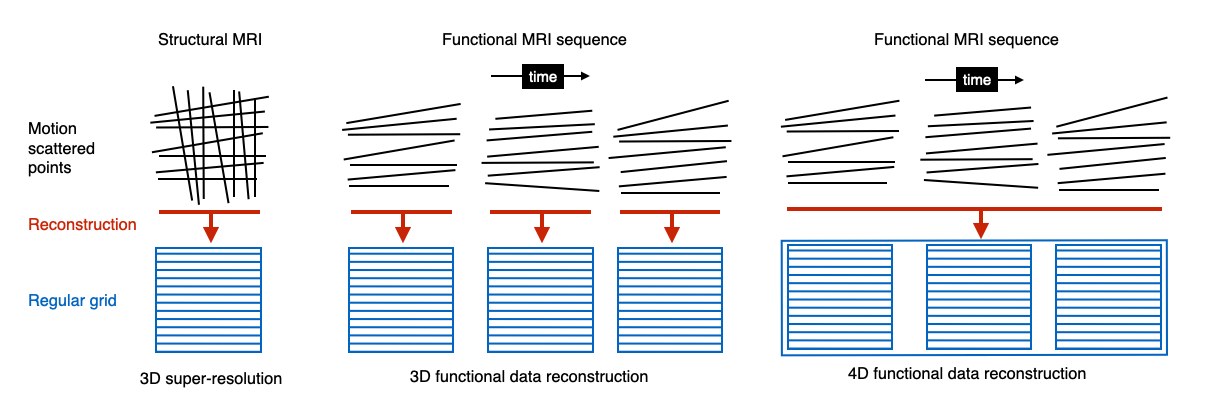}
\caption{Illustration of the image reconstruction using super-resolution technique. Oversampling exists in case of 3D structural MRI (left panel), however, there is not enough data for separate reconstruction of each 3D fMRI volume (middle panel). Here we propose to reconstruct the whole 4D fMRI at once using both spatial and temporal data structure (right panel).} \label{fig1}
\end{figure}

%Thus, we propose to take advantage of the temporal pattern in fMRI to improve reconstruction of fetal \textit{in utero} fMRI data.
Here we propose a new reconstruction method that takes advantage of the temporal structure of fMRI time series and rather than treating each frame independently, it takes both the spatial and the temporal domains into account to iteratively reconstruct a full 4D in utero fMRI image. The proposed method relies on super-resolution techniques that attracted increasing attention in structural fetal T2-weighted imaging, aiming to estimate a 3D high-resolution (HR) volume from multiple (semi-)orthogonal low resolution scans\, \cite{gholipour2010robust,tourbier2015efficient,ebner2020automated}. In case of fMRI, orthogonal acquisitions are not available, instead
%thus there is no spatial oversampling
%but due to the temporal structure of data, neigboring frames can potentially contribute to recover areas with missing data in a single frame.
the reconstruction of a 4D image from a single sequence acquired over time is desired (An illustration of the problem is shown in Figure \ref{fig1}).
Currently, existing single-image reconstruction methods are generally proposed for 3D structural MR images with isotropic voxels, while the effect of motion is implicitly modeled via blurring the desired HR image\,\cite{shi2015lrtv}. None of these methods have been tailored for 4D fMRI with high-levels of movement such as the fetal population. 

Our contribution is threefold: (1) we develop a 4D optimization scheme based on low-rank and total variation regularization to reconstruct 4D fMRI data as a whole (2) we explicitly model the effect of motion in the image degradation process since it is the main source of gaps between interpolated slices; (3) we show the performance of our algorithm on the highly anisotropic \textit{in utero} fMRI images. Experiments were performed on 20 real individuals, and the proposed method was compared to various interpolation methods.  

\section{Method}

We first describe the fMRI image acquisition model and then its corresponding inverse problem formulation to recover a 4D artifact-free fMRI from a single scan of motion corrupted image, using low-rank and total variation regularizations. 

\subsection{The Reconstruction Problem}

fMRI requires the acquisition of a number of volumes over time (fMRI time-series, bold signal) to probe the modulation of spontaneous (or task-related) neural activity. This activity is characterized by low frequency fluctuations ($<0.1 Hz$) of bold signals and therefore temporal smoothing is often applied as a pre-processing step in fMRI analysis. We aim at estimating the motion-compensated reconstruction of fMRI time series ($\mathcal{X} \in \mathbb{R} ^{\hat{B}\times \hat{K}\times \hat{H} \times N}$) from observed motion-contaminated fMRI volumes ($\mathcal{T} \in \mathbb{R} ^{B\times K\times H \times N}$) that integrates temporal smoothing within a full 4D iterative framework. Both $\mathcal{X}$ and $\mathcal{T}$ are composed of $N$ 3D volumes $\mathbf{X}_n,\mathbf{T}_n$ acquired over $N$ timepoints.
%$\left\{T_{n}, \mathbf{X}_{n} \mid n=1,2, \ldots N\right\}$. Note that $T$ is one single acquisition. 
In MR image acquisition, a degradation process yields a low-resolution image from the latent high-resolution image:
\begin{equation} \label{eq:1}
\mathbf{T}_{n} = DSM_{n}\mathbf{X}_{n} + z
\end{equation}
where $D$ is a 3D downsampling operator, $S$ is a 3D blurring operator, $M$ is the set of estimated motion parameters (three rotation and three translation parameters for each slice $\mathbf{t}_{n,h} \in \mathbb{R} ^{\textbf{B}\times K}$ of the volume $\mathbf{T}_{n}$, estimated prior to optimization (Sec.\,\ref{sec:data})), and $z$ represents the observation noise. %The method of estimating $M_{n}$ is explained in the next section and 
The application of $M_{n}$ in the model here is equivalent to transforming each slice by the motion followed by resampling them on a 4D regular grid. 
Successful recovery of $\mathcal{X}$ from the $\mathcal{T}$ not only ensures the compensation of motion but also smoother bold signals due to the implicit temporal structure present in the data. However, since the Eq.(\ref{eq:1}) is ill-posed, direct recovery of $\mathcal{X}$ is not possible without enforcing a prior. Hence, the reconstruction of the latent desired 4D image $\mathcal{X}$ is achieved by minimizing the following cost function based on the inverse problem formulation:
\begin{equation}\label{eq:2}
\min _{\mathcal{X}} \sum_{n=1}^{N}\left\|D S M_{n} \mathbf{X}_{n}-\mathbf{T}_{n}\right\|^{2}+\lambda \Re(\mathcal{X})
\end{equation}
where $\Re(\mathcal{X})$ is a spatio-temporal regularization term, and $\lambda$ balances the contributions of the data fidelity and regularization terms. We propose two regularization terms in this context, 4D low-rank for missing data recovery and total variation for preserving local spatial consistency.

\subsubsection{4D Low-Rank Regularization} Rank as a measure of nondegenerateness of the matrix, is defined by the maximum number of linearly independent rows or columns in the matrix. Since self-similarity is widely observed in fMRI images, low rank prior has been successfully used in matrix completion of censored fMRI time series\,\cite{balachandrasekaran2021reducing}. Here we use low rank as a regularization term to help retrieve relevant information from all image regions. 
To compute the rank for a 4D image $\mathcal{X}$, we first unfold it into a 2D matrix along each dimension\,\cite{liu2012tensor}. Specifically, suppose the size of $\mathcal{X}$ is $B \times K \times H \times N$, we unfold it into four 2D matrices $\left\{X_{(i)}, i=1,2,3,4\right\}$ with size of $B \times\left(K \times H \times N\right), K \times\left(B \times H \times N\right), H \times\left(B \times K \times N\right)$, and $N \times\left(B \times K \times H\right)$ where X(i) means unfold $\mathcal{X}$ along dimension i. Then we compute the sum of the singular values in each matrix for their trace norms $\left\|X_{(i)}\right\|_{t r}$. Finally, the rank of $\mathcal{X}$ is approximated as the combination of trace norms of all unfolded matrices\,\cite{shi2015lrtv}:
\begin{equation}\label{eq:3}
\Re_{r a n k}(\mathcal{X})=\sum_{i=1}^{4} \alpha_{i}\left\|X_{(i)}\right\|_{t r}
\end{equation}
where $\left\{\alpha_{i}\right\}$ are parameters satisfying $\alpha_{i} \geq 0$, and $\sum_{i=1}^{4} \alpha_{i}=1$. By minimizing this term, we obtain a low-rank approximation of $\mathcal{X}$. The low rank regularization is applied in the entire 4D data retrieving useful information for the reconstruction task from both spatial and temporal domains.

\subsubsection{Total Variation Regularization}
Total variation (TV) is defined as integrals of absolute gradient of the signal. For a 4D functional image $\mathcal{X}$:
\begin{equation}\label{eq:4}
\Re_{t v}(\mathcal{X})=\sum_{n=1}^{N} \int\left|\nabla \mathbf{X}_{n}\right| d b d k d h
\end{equation}
where the gradient operator is performed in 3D spatial space. TV regularization has been largely adopted in image recovery because of its powerful ability in edge preservation\,\cite{tourbier2015efficient,shi2015lrtv}. Here, we use TV in 3D space instead of 4D space based on the notion that primarily the spatial neighborhood exhibits consistency %across different functional volumes 
and thus TV in temporal domain may not be effective. 
\subsection{Optimization} The proposed 4D single acquisition reconstruction is thus formulated as below:
\begin{equation}\label{eq:5}
\min _{\mathcal{X}} \sum_{n=1}^{N}\left\|D S M_{n} \mathbf{X}_{n}-\mathbf{T}_{n}\right\|^{2}+\lambda_{\text {rank }} \Re_{\text {rank }}(\mathcal{X})+\lambda_{t v} \sum_{n=1}^{N} \Re_{t v}\left(\mathbf{X}_{n}\right)
\end{equation}
We employ the alternating direction method of multipliers (ADMM) algorithm to minimize the cost function in Eq.(\ref{eq:5}). ADMM has been proven efficient for solving optimization problems with multiple non-smooth terms\,\cite{boyd2011distributed}. Briefly, we first introduce redundant variables $\left\{Y_{i}\right\}_{i=1}^{4}$ with equality constraints $
\mathcal{X}_{(i)}=Y_{i(i)}$, and then use Lagrangian dual variables $\left\{U_{i}\right\}_{i=1}^{4}$ to integrate the equality constraints into the cost function:
\begin{equation}
\begin{array}{l}
\min _{{\mathcal{X}},\left\{Y_{i}\right\}_{i=1}^{4},\left\{U_{i}\right\}_{i=1}^{4}} \sum_{n=1}^{N}\left\|D S M_{n} \mathbf{X}_{n}-\mathbf{T}_{n}\right\|^{2}+\lambda_{\text {rank }} \sum_{i=1}^{4} \alpha_{i}\left\|Y_{i(i)}\right\|_{t r} \\
{\quad+\sum_{i=1}^{4} \frac{\rho}{2}\left(\left\|\mathcal{X}-Y_{i}+U_{i}\right\|^{2}-\left\|U_{i}\right\|^{2}\right)+\lambda_{t v} \sum_{n=1}^{N} \int\left|\nabla \mathbf{X}_{n}\right| d b d k d h}
\end{array}
\end{equation}
We break the cost function into subproblems for $\mathcal{X}$, Y, and U, and iteratively update them. The optimization scheme is summarized in Algorithm \ref{alg:cap}. 
\renewcommand{\algorithmicrequire}{ \textbf{Input:}}
\renewcommand{\algorithmicensure}{ \textbf{Initialize:}}
\begin{algorithm}[t]
\caption{4D motion-compensated reconstruction of fMRI time series}\label{alg:cap}
\begin{algorithmic}
\Require{Single scan fMRI image $\mathcal{T}$, realignment parameters}
\Ensure{The desired $\mathcal{X}$ by resampling motion-transformed image $\mathcal{T}$ with linear interpolation. Set auxiliary variable $Y_{i}^{(0)}=0, U_{i}^{(0)}=0, i=1,2,3,4$}
\While{$\left\|\mathcal{X}^{k}-\mathcal{X}^{k-1}\right\| /\|\mathcal{T}\| > \varepsilon$}
\State{Update $\mathcal{X}^{k}$ by using gradient descent:}
\State $
\arg \min _{\mathcal{X}} \sum_{n=1}^{N}\left\|D S M_{n} \mathbf{X}_{n}^{(k-1)}-\mathbf{T}_{n}\right\|^{2}+\sum_{i=1}^{4} \frac{\rho}{2}\left\|\mathcal{X}^{(k-1)}-Y_{i}^{(k-1)}+U_{i}^{(k-1)}\right\|^{2} + \lambda_{t v} \sum_{n=1}^{N} \int\left|\nabla \mathbf{X}_{n}^{(k-1)}\right| d b d h d k
\hfill\inlineeqno$
\State{Update $Y_{i}^{(k)}$ by using Singular Value Thresholding:}
\State{$Y_{i}^{(k)}=fold_{i}\left[SVT_{\lambda_{rank }\alpha_{i} / \rho}\left(\mathcal{X}_{(i)}^{(k)}+U_{i(i)}^{(k-1)}\right)\right] \hfill\inlineeqno$}
\State{with $fold_{i}\left(Y_{i(i)}\right)=Y_{i}$}
\State{Update $U_{i}^{(k)}=U_{i}^{(k-1)}+\left(\mathcal{X}^{(k)}-Y_{i}^{(k)}\right) \hfill\inlineeqno$}
\EndWhile
\end{algorithmic}
\end{algorithm}

\section{Experiments and Results}
\subsection{Data}\label{sec:data}
\textbf{\textit{Data acquisition:}} Experiments in this study were performed on 20 \textit{in utero} fMRI sequences obtained from fetuses between 19 and 39 weeks of gestation. None of the cases showed any neurological pathology. Pregnant women were scanned on a 1.5T clinical scanner (Philips Medical Systems, Best, Netherlands) using single-shot echo-planar imaging (EPI), and a sensitivity encoding (SENSE) cardiac coil with five elements. Image matrix size was 144$\times$144, with 1.74$\times$1.74$mm^{2}$ in-plane resolution, 3$mm$ slice thickness, a TR/TE of 1000/50 ms, and a flip angle of 90\degree. Each scan contains 96 volumes obtained in an interleaved slice order to minimize cross-talk between adjacent slices.

\noindent \textbf{\textit{Preprocessing:}} For preprocessing, a binary brain mask was manually delineated on the average volume of each fetus and dilated to ensure it covered the fetal brain through all ranges of the motion. A four dimensional estimate of the bias field for spatio-temporal signal non-uniformity correction in fMRI series was obtained using N4ITK algorithm\,\cite{tustison2010n4itk} as suggested previously\,\cite{seshamani2014method}. Intensity normalization was performed as implemented in mialSRTK toolkit\,\cite{tourbier2017automated}. Finally, motion parameters were estimated by performing a hierarchical slice-to-volume registration based on the interleaved factor of acquisition to a target volume created by automatically finding a set of consecutive volumes of fetal quiescence and averaging over them\,\cite{seshamani2013cascaded}. Image registration software package NiftyReg\,\cite{modat2014global} was used for all motion correction steps in our approach. Demographic information of all 20 subjects as well as the maximum motion parameters estimated were reported in Supplementary Table S1.
%~\ref{tab:vae}.

\begin{figure}[t!]
\includegraphics[width=\textwidth]{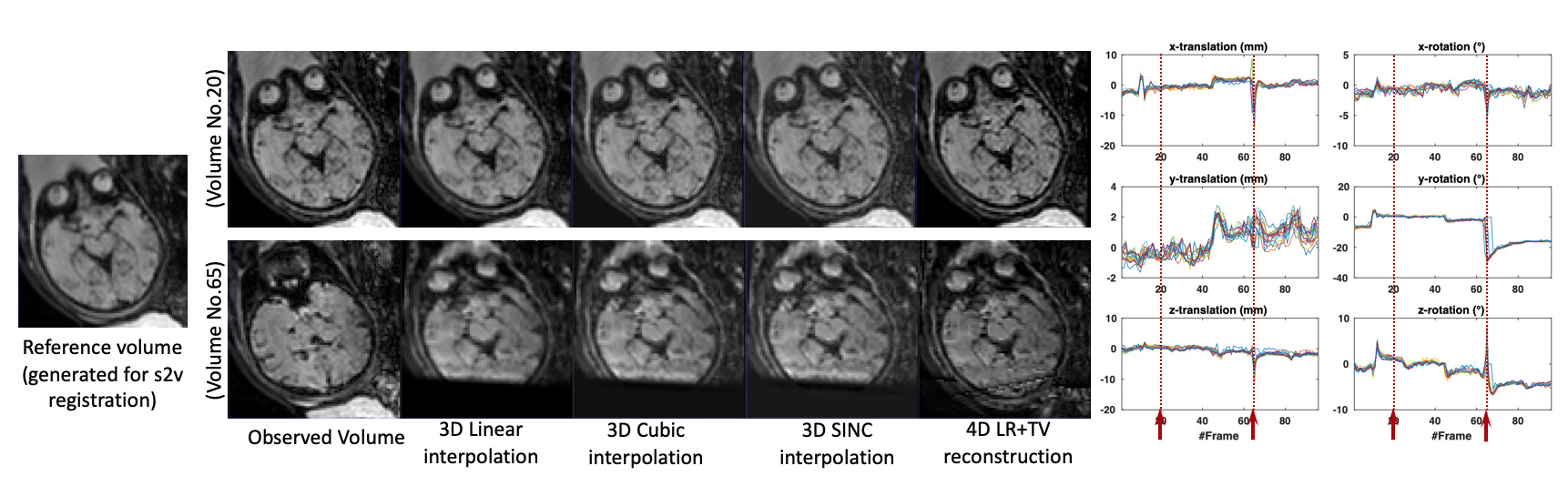}
\caption{Reconstruction of \textit{in-utero} fMRI for a typical fetus, and the estimated slice-wise realignment parameters. When motion is small (volume No.20) all interpolation methods recovered a motion compensated volume, and our approach resulted in a sharper image. In contrast, with strong motion relative to the reference volume (volume No.65), single step 3D interpolation methods are not able to recover the whole brain, and parts remain missing, whereas the proposed 4D iterative reconstruction did recover the entire brain.} \label{fig2}
\end{figure}
\subsection{Experimental Setting and Low-Rank Representation}
\sloppy We first evaluated to which extent \textit{in utero} fMRI data can be characterized by its low-rank decomposition. The rapid decay of the singular values for a representative slice of our cohort is shown in Supplementary Figure S1.
%~\ref{figS1}. 
We used the top 30, 60, 90, and 120 singular values to reconstruct this slice and measured signal-to-noise ratio (SNR)
%as $SNR=20\log 10(\|f\| /\|f-g\|)$ 
to evaluate the reconstruction accuracy. The number of used singular values determines the rank of the reconstructed image. Using the top 90 or 120 singular values (out of 144), the reconstructed image does not show visual differences compared to the original image while it has a relatively high SNR (Figure S1).   
%~\ref{figS1}

For the full 4D fMRI data of our cohort with the size of 144$\times$144$\times$18$\times$96, four ranks, one for each unfolded matrix along one dimension is computed. Each is less than the largest image size 144. These ranks are relatively low in comparison to the total number of elements, implying \textit{in utero} fMRI images could be represented using their low-rank approximations. We set $\alpha_{1}=\alpha_{2}=\alpha_{3}=\alpha_{4}=1/4$ as all dimensions are assumed to be equally important, $\lambda_{\text {rank}}=0.01$, $\lambda_{\text {tv}}=0.01$ were chosen empirically. The algorithm stopped when the difference in iterations was less than $\varepsilon=1 e-5$.
\begin{figure}[!b]
\includegraphics[width=\textwidth]{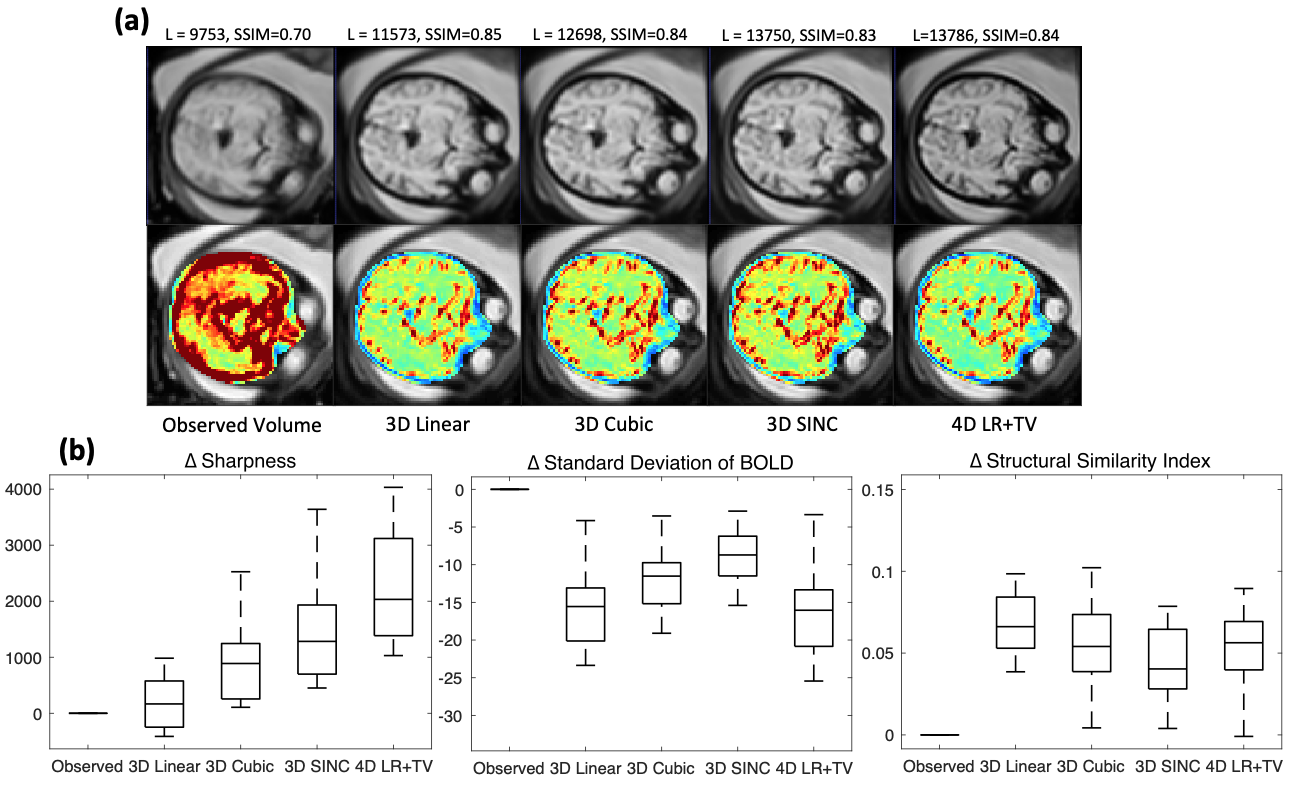}
\caption{Evaluation metrics for a typical fetus (a) and the whole cohort (b). Panel (a) shows an example slice in the average volume (top row) and voxel-wise standard deviation of the bold signal during fMRI acquisition. Higher Laplacian (sharpness) and SSIM, and lower standard deviation are indicative of better recovery. Panel (b) demonstrates these metrics in our fetal dataset.} \label{fig3}
\end{figure}
\subsection{Evaluation of Image Reconstruction} 
A number of interpolation methods was employed to be compared with our reconstructed image including linear, cubic spline, and SINC interpolation. For each method, we applied the same realignment parameters as the ones used in our model, and in accordance with standard motion correction techniques, each 3D volumes of fMRI time series was interpolated separately. We quantified sharpness\,\cite{pech2000diatom} of the average recovered image, standard deviation of bold signal fluctuations\,(SD) through-out the sequence, and the Structural Similarity Index\,(SSIM) which correlates with the quality of human visual perception\,\cite{wang2004image}. Higher values of sharpness and SSIM, and lower values of SD are indicative of better recovery. 
%Note that in our real application of \textit{in utero} imaging, there is no ground truth for fully motion compensated image and also the synthetic data without motion does not represent the problem we want to address in this paper since the challenge comes from areas with missing data after volumetric scatter data interpolation. Therefore, we could only compare different methods of image recovery without having a real ground truth as shown for a typical fetus in Fig.\,\ref{fig2}. 

%From left to right, the first column of 
Figure\,\ref{fig2} shows, from left to right, the reference volume, two corresponding slices in the observed image, and the results of different reconstruction methods. Volume No.20 exhibits minor motion, volume No.65 exhibits strong motion. The motion estimate plots on the right show their respective time points.
%shows the corresponding observed slices in two different volumes- one with major rotation (volume No.65) and one with minor rotation (volume No.20). 
%These volumes were chosen according to the realignment parameters estimated with registration of each slice to a generated reference volume. 
The figure shows the recovered slices of these two volumes using 3D linear, cubic, SINC, and the proposed 4D LR+TV method, respectively. In the case of excessive complex motion (30$\degree$ out of the plane combined with in-plane rotation and translation), the 3D interpolation methods cannot recover the whole slice as they utilize information only from the local spatial neighborhoods. The reconstructed slice by the proposed 4D iterative reconstruction approach recovers the image information, is sharper, and preserves more structural detail of the brain. 
Figure\,\ref{fig3} shows a qualitative and quantitative comparison of reconstruction approaches. Figure\,\ref{fig3}\,(a) 
%the overall performance of the algorithm across the whole volumes of fMRI data, 
shows the average volume (top row), and the standard deviation of intensity changes over time (bottom row) for one subject. 4D reconstruction achieves sharper structural detail, and overall reduction of the standard deviation, which is primarily related to motion as described earlier. 
%As mentioned earlier, smooth bold signal along time dimension and therefore lower standard deviation is desired. 
Although linear interpolation results in signals as smooth as the proposed method, severe blurring is observed in the obtained image by this approach. 
%
% AVERAGE OVER THE POPULATION
% sharpness: (0    0.2946    0.9593    1.5211    2.2938) * 1000
% SD: 69.44, 52.94, 56.74, 60.10, and 52.43
% SSIM: 0.7987, 0.8705, 0.8573, 0.8465, 0.8566
% AVERAGE IMPROVEMENT COMPARED TO OBSERVED: <- i think we should put in these values
% delta Sharpness (0 0.2946 0.9593 1.5211 2.2938)*1000
% delat std:  0  -16.4981  -12.6945   -9.3392  -17.0097
%
Figure\,\ref{fig3}\,(b) provides the quantitative evaluation for the entire study population. The proposed method significantly (p$<$0.01, paired-sample t-tests for each comparison) outperforms all comparison methods. The average gain of sharpness over the observed image is 2294 in our method compared to 1521 for 3D SINC, 959 for 3D Cubic, and 294 for 3D Linear, and the average reduction of SD relative to the observed image is -17 in our method compared to -9.34 for 3D SINC, -12.70 for 3D Cubic, and -16.50 for 3D Linear. The difference between linear interpolation and our approach did not reach the statistical significance level for SSIM (p=0.28). In summary, 4D iterative reconstruction reduces standard deviation over time, while increasing sharpness and recovered structure, which the 3D approaches failed to achieve. 
%The proposed algorithm not only reduces signal standard deviation over time but but also detailed structure and enhanced sharpness.

\subsection{Functional Connectivity Analysis}
\begin{figure}[!t]
\includegraphics[width=\textwidth]{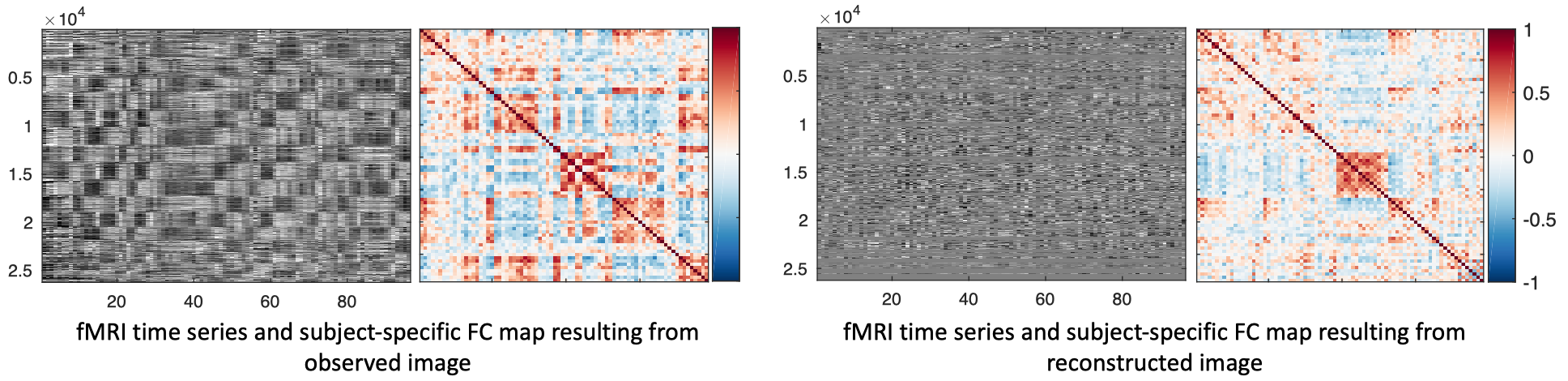}
\caption{Carpet plot and functional connectivity maps achieved for an example subject using the observed fMRI time series and the time series recovered by 4D iterative reconstruction.} \label{fig5}
\end{figure}

Figure\,\ref{fig5} illustrates the impact of the accurate motion correction and reconstruction for the analysis of functional connectivity (FC) in the fetal population. The details of the pipeline employed for extracting subject-specific FC maps is explained in the supplementary material. When using the time series recovered by our proposed approach for FC analysis, the number of motion-corrupted correlations decreased significantly as visible in the \textit{carpet plot} of signals, and the associated connectivity matrix. 
%This finding was not apparent when applying standard approach.

\section{Conclusion}

In this work, we presented a novel spatio-temporal iterative 4D reconstruction approach for \textit{in-utero} fMRI acquired while there is unconstrained motion of the head. The approach utilizes the self-similarity of fMRI data in the temporal domain as 4D low-rank regularisation together with total variation regularization based on spatial coherency of neighboring voxels. Comparative evaluations on 20 fetuses show that this approach yields a 4D signal with low motion induced standard deviation, and recovery of fine structural detail, outperforming various 3D reconstruction approaches. 
%. Linear, Spline, and Sinc interpolation showed inferior performance, which may be because the image degradation is not explicitly modeled in these methods.

\section*{Acknowledgment}
%\label{sec:majhead}
This work has received funding from the European Union's Horizon 2020 research and innovation programme under the Marie Sk\l odowska-Curie grant agreement No 765148, and partial funding from the Austrian Science Fund (FWF, P35189, I 3925-B27), in collaboration with the French National Research Agency (ANR), and the Vienna Science and Technology Fund (WWTF, LS20-065). This work was also supported by the Swiss National Science Foundation (project 205321-182602). We acknowledge access to the expertise of the CIBM Center for Biomedical Imaging, a Swiss research center of excellence founded and supported by Lausanne University Hospital (CHUV), University of Lausanne (UNIL), Ecole polytechnique fédérale de Lausanne (EPFL), University of Geneva (UNIGE) and Geneva University Hospitals (HUG). 

%This work was supported by the Medi- cal University of Vienna, the Austrian Research Fund (FWF) [grants P 35189, P 34198, and I 3925-B27] in collaboration with the French National Research Agency (ANR), the Vienna Science and Technol- ogy Fund (WWTF) [LS20-065], 


\bibliographystyle{splncs04}
\bibliography{paper1715.bib}
%\end{thebibliography}
\afterpage{
\beginsupplement
\begin{landscape}
\centering
\begin{table}
\small
{\caption{Gestational age, motion characteristics, and the achieved values of the evaluation metrics including sharpness (quantified by Laplacian), standard deviation of BOLD signal fluctuations (SD), and structural similarity index (SSIM) of all fetuses. The proposed method (LRTV) outperformed other reconstruction methods in terms of sharpness and SD significantly, and the difference between linear interpolation and our approach did not reach the statistical significance level for SSIM (p=0.28)}\label{tab:vae}}
\resizebox{\linewidth}{!}{
%\begin{tabularx}{1.5\textwidth}{l c *{6}{l} *{12}{X}}
\begin{tabular}{l c l l l l l l || c c c c c | c c c c c | c c c c c}
\toprule
    &  \textbf{GA} & \multicolumn{3}{c}{\textbf{max trans(\small{mm})}} & \multicolumn{3}{c||}{\textbf{max rot(\textbf{\degree)}}} &\multicolumn{5}{c|}{\textbf{Sharpness}} &\multicolumn{5}{c|}{\textbf{SD}}
    &\multicolumn{5}{c}{\textbf{SSIM}}\\
    &  \textbf{(week+day)} & \textbf{x}& \textbf{y}& \textbf{z} & \textbf{roll} & \textbf{pitch} & \textbf{yaw} &  \textbf{Raw} & \textbf{Linear} & \textbf{Cubic} & \textbf{Sinc} & \textbf{LRTV} & \textbf{Raw} & \textbf{Linear} & \textbf{Cubic} & \textbf{Sinc} & \textbf{LRTV} & \textbf{Raw} & \textbf{Linear} & \textbf{Cubic} & \textbf{Sinc} & \textbf{LRTV}\\ 
\midrule
    \textbf{S1} & 19w5d &6.87 & 8.06 &	2.89 & 6.17 & 5.86 & 4.48 &  1159 & 1145 & 12504 & 13161 & \textbf{13786} & 38.29 & \textbf{34.15} & 34.76 & 35.40 & 34.94 & 0.9682 & \textbf{0.9727} & 0.9725 & 0.9721 & 0.9673 \\ 
     \textbf{S2} & 22w5d &21.86 & 8.11 & 19.29 & 27.88 & 20.63 & 27.95 & 9383 & 9523 & 10187 & 10721 & \textbf{11053} & 77.43 & 59.11 & 64.36 & 68.70 & \textbf{58.26} & 0.8755 & \textbf{0.9210} & 0.9109 & 0.9031 & 0.9105 \\ 
    \textbf{S3}  & 28w3d &2.04 & 1.94 & 13.73 & 10.27 & 17.16 & 1.51 &  6610 & 7352 & 9094 & 10251 & \textbf{10282} & 59.85 & 46.05 & 50.09 & 53.63 & \textbf{45.82} & 0.8229 & \textbf{0.8807} & 0.8684 & 0.8577 & 0.8673 \\ 
    \textbf{S4}  & 29w2d &2.39 & 4.00 & 6.95 & 5.18 & 3.82 & 2.81& 10636 & 10435 & 10776 & 11214 & \textbf{11864} & 59.38 & \textbf{45.45} & 50.31 & 54.59 & 45.95 & 0.8046 & \textbf{0.8650} & 0.8465 & 0.8323 & 0.8499 \\ 
    \textbf{S5}  & 27w1d &7.34 & 3.93 & 4.61 & 1.66 & 4.09 & 2.79& 6533 & 9049 & 9059 & 9159 & \textbf{9314} & 59.22 & 46.85 & 49.51 & 52.00 & \textbf{45.98} & 0.8216 & \textbf{0.8885} & 0.8825 & 0.8772 & 0.8790 \\ 
    \textbf{S6}  & 25w5d  &13.83 & 5.56 & 7.84 & 13.88 & 25.73 & 15.54 & 9134 & 9379 & 9977 & 10428 & \textbf{10979} & 77.49 & 56.74 & 61.79 & 65.98 & \textbf{55.17} & 0.8125 & \textbf{0.8758} & 0.8609 & 0.8503 & 0.8679 \\ 
    \textbf{S7}  & 26w3d & 5.97 & 2.42 & 2.37 & 2.23 & 3.54 & 2.60 & 9168 & 8928 & 9462 & 9919 & \textbf{10405} & 57.61 & 49.71 & 52.14 & 54.56 & \textbf{48.87} & 0.8519 & \textbf{0.8905} & 0.8825 & 0.8742 & 0.8787 \\ 
    \textbf{S8}  & 30w2d  & 3.94 & 7.45 & 3.93 & 16.63 & 2.34 & 21.84 & 8942 & 8530 & 9156 & 9780 & \textbf{10245} & 75.52 & \textbf{56.02} & 59.96 & 63.29 & 56.57 & 0.8241 & \textbf{0.8895} & 0.8789 & 0.8703 & 0.8814 \\ 
    \textbf{S9}  & 29w5d  & 5.23 & 6.99 & 2.24 & 6.55 & 5.29 & 4.95 & 9753 & 11573 & 12698 & 113751 & \textbf{13786} & 76.71 & \textbf{53.89} & 58.10 & 61.90 & 53.97 & 0.7734 & \textbf{0.8618} & 0.8517 & 0.8432 & 0.8487 \\ 
    \textbf{S10}  & 31w6d  & 11.12 & 3.68 & 3.97 & 3.06 & 13.96 & 4.16 & 9223 & 9658 & 10368 & 11074 & \textbf{11444} & 63.67 & 48.41 & 53.08 & 57.19 & \textbf{48.05} & 0.7821 & \textbf{0.8302} & 0.8139 & 0.7999 & 0.8132 \\ 
    \textbf{S11}  & 32w4d  & 3.80 & 2.00 & 2.09 & 2.87 & 2.49 & 1.75 & 8715 & 8459 & 9205 & 9817 & \textbf{9754} & 69.73 & 53.88 & 57.32 & 60.43 & \textbf{53.78} & 0.7582 & \textbf{0.8382} & 0.8255 & 0.8149 & 0.8194 \\ 
    \textbf{S12}  & 29w5d  & 3.48 & 3.35 & 3.33 & 2.00 & 2.19 & 4.69 & 16344 & 16142 & 17332 & 18436 & \textbf{19875} & 72.10 & 55.71 & 61.49 & 66.50 & \textbf{55.25} & 0.7776 & \textbf{0.8570} & 0.8368 & 0.8205 & 0.8389\\ 
    \textbf{S13}  & 36w1d  & 2.17 & 2.75 & 3.80 & 9.54 & 5.96 & 6.22 & 9862 & 9531 & 10088 & 10509 & \textbf{11122} & 87.33 & 54.95 & 58.48 & 61.65 & \textbf{54.28} & 0.7041 & \textbf{0.8554} & 0.8422 & 0.8302 & 0.8390 \\ 
    \textbf{S14}  & 34w5d  & 42.33 & 20.62 & 15.89 & 7.33 & 16.43 & 10.52 & 9230 & 9051 & 10164 & 11093 & \textbf{12054} & 64.13 & 45.16 & 49.59 & 53.43 & \textbf{44.74} & 0.6508 & \textbf{0.7839} & 0.7529 & 0.7293 & 0.7655 \\ 
    \textbf{S15}  & 23w6d  & 12.93 & 12.44 & 7.42 & 3.04 & 4.74 & 4.29 & 20016 & 20999 & 21363 & 21625 & \textbf{23388} & 70.93 & 55.90 & 60.42 & 64.48 & \textbf{55.21} & 0.8264 & \textbf{0.8904} & 0.8746 & 0.8615 & 0.8756 \\ 
    \textbf{S16}  & 29w4d  & 2.83 & 5.09 & 3.10 & 19.82 & 12.99 & 21.57 & 19293 & 19943 & 20304 & 20566 & \textbf{22160} & 77.39 & 54.03 & 58.29 & 61.99 & \textbf{51.95} & 0.7637 & \textbf{0.8617} & 0.8444 & 0.8314 & 0.8531 \\ 
    \textbf{S17}  & 29w3d  & 20.97 & 9.18 & 8.07 & 7.61 & 17.55 & 16.10 & 11628 & 11225 & 11735 & 12266 & \textbf{13096} & 79.02 & 57.64 & 64.22 & 69.69 & \textbf{56.74} & 0.7582 & \textbf{0.8331} & 0.8114 & 0.7955 & 0.8215 \\ 
    \textbf{S18}  & 24w4d  & 4.33 & 5.65 & 6.13 & 6.05 & 3.97 & 5.04 & 11628 & 11225 & 11735 & 12266 & \textbf{13096} & 64.57 & 53.39 & 56.06 & 58.36 & \textbf{52.26} & 0.8657 & \textbf{0.9084} & 0.9005 & 0.8942 & 0.8955\\ 
    \textbf{S19}  & 34w3d  & 4.48 & 1.51 & 12.55 & 10.16 & 12.93 & 3.16 & 8454 & 8959 & 9404 & 9651 & \textbf{10116} & 88.02 & 72.88 & 74.73 & 76.55 & \textbf{71.88} & 0.7576 & \textbf{0.8561} & 0.8443 & 0.8344 & 0.8341 \\ 
    \textbf{S20}  & 39w2d & 0.94 & 3.78 & 1.06 & 3.57 & 2.95 & 1.74 & 17439 & 17633 & 17725 & 17891 & \textbf{21223} & 70.36 & \textbf{58.90} & 60.18 & 61.68 & 58.92 & 0.7554 & \textbf{0.8494} & 0.8443 & 0.8367 & 0.8256\\ 
    \bottomrule
    \end{tabular}}
\end{table}
\end{landscape}
}
\newpage
\beginsupplement
\begin{figure}[!t]
\centering
\includegraphics[width=0.8\textwidth]{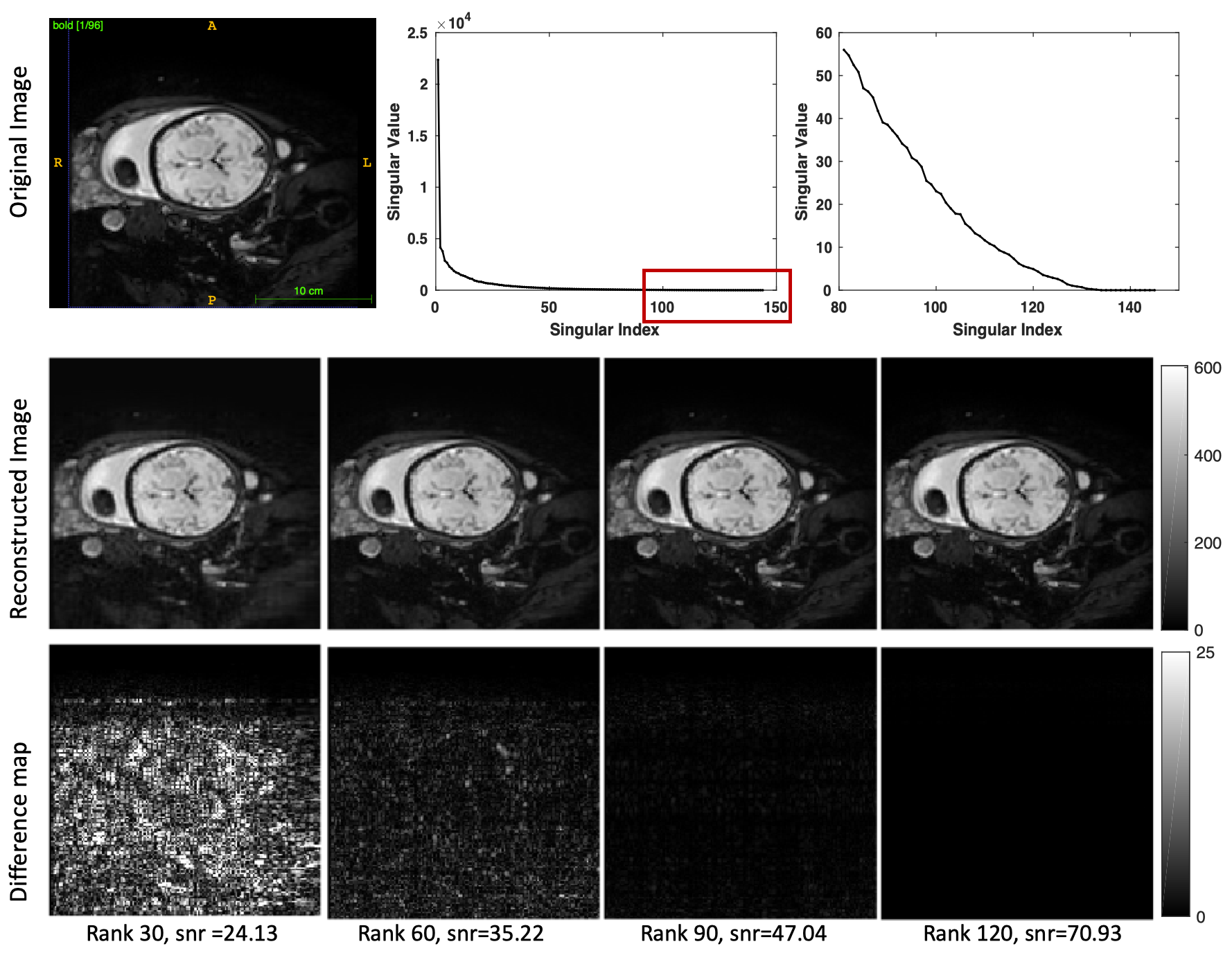}
\caption{Low rank approximation of \textit{in-utero} fMRI of a fetus with gestational age of 34w+4d. Top row shows the original slice, singular-value plot, and zoomed singular-value plot of indices from 80 to 144. Bottom row shows the four reconstructed slices and their differences with the original image by using top 30, 60, 90, and 120 singular values, respectively. SNR values were reported at the bottom of each reconstructed slices.} \label{figS1}
\end{figure}
\begin{figure}
\centering
\includegraphics[width=\textwidth]{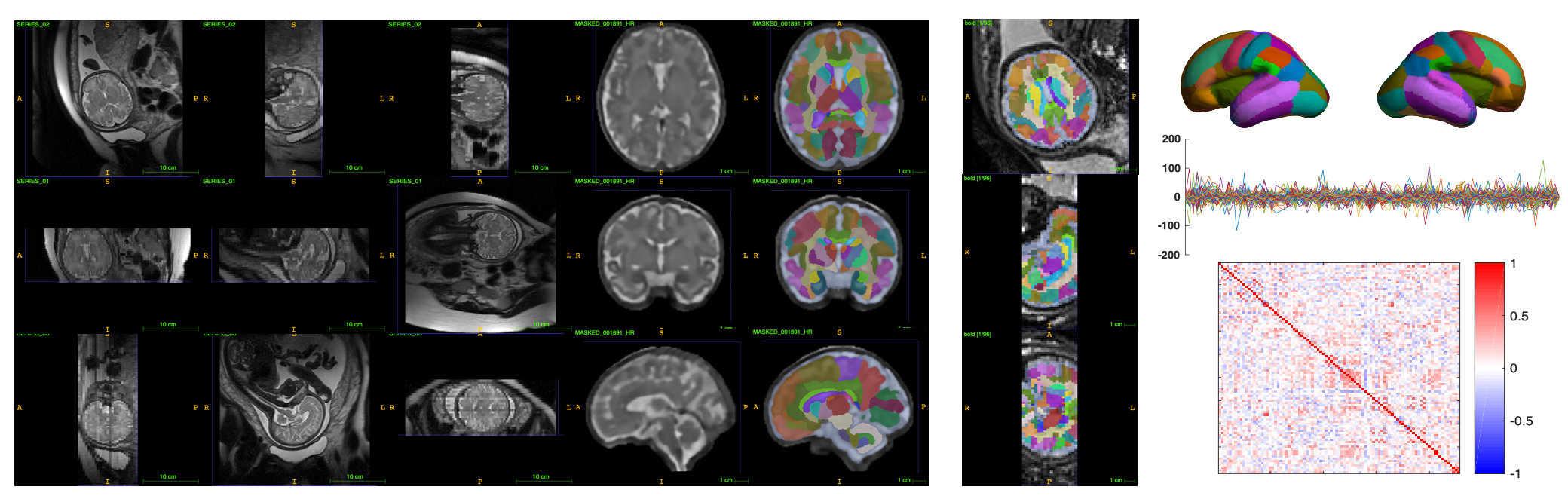}
\caption{subject-specific functional connectivity analysis was performed in the native functional space. For this, cortical ROIs were first obtained using an automatic atlas-based segmentation of $T2$ scans acquired during the same session as the fMRI, using a publicly available atlas of fetal brain anatomy \cite{gholipour2017normative}. The resulting parcellation consists of 78 ROIs and was mapped to the motion corrected fMRI space using a rigid transformation. For each parcel, the average time series of all voxels was computed, and aCompCor nuisance regression and temporal filtering were performed subsequently. FC matrix was estimated by measuring Pearson's correlation between the average time series of parcels.} \label{figS2}
\end{figure}
\end{document}